\documentclass[sigconf]{acmart}
\AtBeginDocument{%
  }

\copyrightyear{2026}
\acmYear{2026}
\setcopyright{cc}
\setcctype{by}
\acmConference[WSDM '26]{Proceedings of the Nineteenth ACM International Conference on Web Search and Data Mining}{February 22--26, 2026}{Boise, ID, USA}
\acmBooktitle{Proceedings of the Nineteenth ACM International Conference on Web Search and Data Mining (WSDM '26), February 22--26, 2026, Boise, ID, USA}
\acmPrice{}
\acmDOI{10.1145/3773966.3779399}
\acmISBN{979-8-4007-2292-9/2026/02}

\usepackage[table]{xcolor}
\usepackage[most]{tcolorbox}
\usepackage{pgfplots}
\usetikzlibrary{pgfplots.groupplots}
\usepackage{subcaption}
\usepackage{rotating}
\usepackage{multirow}
\usepackage{makecell}
\usepackage{graphicx}
\usepackage{arydshln}
\pgfplotsset{compat=1.18}
\usepackage{balance}

\definecolor{zhucao}{rgb}{0.65, 0.25, 0.21}
\definecolor{fuguang}{rgb}{0.75, 0.57, 0.45}   
\definecolor{tushoulan}{rgb}{0.30, 0.56, 0.70} 
\definecolor{qingdai}{rgb}{0.31, 0.52, 0.41}   
\definecolor{bglight}{RGB}{230, 230, 230}
\definecolor{lightyellow}{RGB}{245, 245, 220}   
\definecolor{lightblue}{RGB}{224, 240, 255}   

\begin{document}

\title[Accelerating Generative Recommendation via Simple Categorical User Sequence Compression]{Accelerating Generative Recommendation via Simple Categorical User Sequence Compression
}

\author{Qijiong Liu}
\authornote{Equal contribution (co-first authors).}
\affiliation{%
  \institution{The HK PolyU}
  \country{Hong Kong SAR}
}

\author{Lu Fan}
\authornotemark[1]
\affiliation{%
  \institution{The HK PolyU}
  \country{Hong Kong SAR}
}

\author{Zhongzhou Liu}
\affiliation{%
  \institution{Huawei}
  \country{Shenzhen, China}
}

\author{Xiaoyu Dong}
\affiliation{%
  \institution{The HK PolyU}
  \country{Hong Kong SAR}
}

\author{Yuankai	Luo}
\affiliation{%
  \institution{The HK PolyU}
  \country{Hong Kong SAR}
}

\author{Guoyuan	An}
\affiliation{%
  \institution{Huawei}
  \country{Shenzhen, China}
}

\author{Nuo	Chen}
\affiliation{%
  \institution{The HK PolyU}
  \country{Hong Kong SAR}
}

\author{Wei	Guo}
\affiliation{%
  \institution{Huawei}
  \country{Shenzhen, China}
}

\author{Yong Liu}
\affiliation{%
  \institution{Huawei}
  \country{Shenzhen, China}
}

\author{Xiao-Ming Wu}
\authornote{Xiao-Ming Wu is the corresponding author. ``The HK PolyU'' refers to ``The Hong Kong Polytechnic University, and ``Huawei'' refers to ``Huawei Technologies Co., Ltd.''. Correspondence: 
Qijiong Liu (\texttt{liu@qijiong.work}), 
Yong Liu (\texttt{liu.yong6@huawei.com}), and 
Xiao-Ming Wu (\texttt{xiao-ming.wu@polyu.edu.hk}).}
\affiliation{%
  \institution{The HK PolyU}
  \country{Hong Kong SAR}
}

\renewcommand{\shortauthors}{Qijiong Liu et al.}
\newcommand{\model}{\texttt{CAUSE}}

\newcommand{\textbox}[1]{\vspace{.5em}\begin{tcolorbox}[
  colback=gray!5!white,           
  colframe=gray!70!black,         
  fonttitle=\bfseries,            
  boxrule=1pt,                    
  arc=4pt,                        
  outer arc=4pt,                  
  top=4pt,                        
  bottom=4pt,                     
  left=4pt,
  right=4pt,
  before skip=10pt,               
  after skip=10pt,                
]
#1
\end{tcolorbox}}

\begin{abstract}
Although generative recommenders demonstrate improved performance with longer sequences, their real-time deployment is hindered by substantial computational costs. To address this challenge, we propose a simple yet effective method for compressing long-term user histories by leveraging inherent item categorical features, thereby preserving user interests while enhancing efficiency. Experiments on two large-scale datasets demonstrate that, compared to the influential HSTU model, our approach achieves up to a 6× reduction in computational cost and up to 39\% higher accuracy at comparable cost (i.e., similar sequence length). 
The source code will be available at \url{https://github.com/Genemmender/CAUSE}.

\end{abstract}



\maketitle

\section{Introduction}

Generative recommendation (GR)~\cite{rajput2023recommender, zheng2023adapting,hstu,g1,g2,g3,liu2024store,zhou2025recbase} is an emerging paradigm for sequential recommendation~\cite{hidasi2016sessionbasedrecommendationsrecurrentneural,boka2024survey}, which directly generate recommendation items rather than using conventional multi-stage recommendation pipeline.
Driven by the \textbf{scaling laws} uncovered in large language models~\cite{kaplan2020scaling,achiam2023gpt}, GR models have recently seen a wave of innovations guided by these principles.

\begin{figure}
\centering
\setlength\tabcolsep{0pt}

\resizebox{.7\linewidth}{!}{
\begin{tikzpicture}
  \begin{axis}[
    width=12cm,
    height=6cm,
    xlabel={Training Time (ms)},
    ylabel={Performance (nDCG@10)},
    grid=both,
    grid style={dashed,gray!30},
    legend style={font=\normalsize,legend cell align=left},
    legend pos=south east,
    tick style={semithick},
    mark size=2.5pt,
  ]

  \addplot[color=zhucao, only marks, style={mark=*, fill=zhucao,mark size=3pt}] coordinates {(49414,0.0173)};
  \label{plot:hstu-l64}

  \addplot[color=tushoulan, only marks, style={mark=*, fill=tushoulan,mark size=3pt}] coordinates {(79040,0.0175)};
  \label{plot:hstu-l512}

  \addplot[color=fuguang, only marks, style={mark=*, fill=fuguang,mark size=3pt}] coordinates {(131800,0.0178)};
  \label{plot:hstu-l1024}

  \addplot[color=qingdai, only marks, style={mark=*, fill=qingdai,mark size=3pt}] coordinates {(297345,0.0183)};
  \label{plot:hstu-l2048}

  \addplot[color=black, thick, dashed, smooth] 
    coordinates {(49414,0.0173) (79040,0.0175) (131800,0.0178) (297345,0.0183)};

  \addplot[color=zhucao, only marks, style={mark=diamond*, fill=zhucao,mark size=4pt}] coordinates {(49414,0.0177)};
  \label{plot:usc-l64}

  \addplot[color=tushoulan, only marks, style={mark=diamond*, fill=tushoulan,mark size=4pt}] coordinates {(79040,0.0188)};
  \label{plot:usc-l512}

  \addplot[color=black, thick, dashed, smooth] 
    coordinates {(49414,0.0177) (79040,0.0188)};

  \end{axis}

\node[draw,fill=white,inner sep=4pt,above left=0.5em] at (10.3, 0) {
    \setlength\tabcolsep{2pt}
    \resizebox{.25\linewidth}{!}{
    \begin{tabular}{ccl}
    $\texttt{HSTU}$ & $\textit{ours}$ & \texttt{Length} \\
    \ref{plot:hstu-l64} & \ref{plot:usc-l64} & $\texttt{64}$\\
    \ref{plot:hstu-l512} & \ref{plot:usc-l512} & $\texttt{512}$\\
    \ref{plot:hstu-l1024} &  & $\texttt{1,024}$ \\
    \ref{plot:hstu-l2048} &  & $\texttt{2,048}$
    \end{tabular}
    }
    };
\end{tikzpicture}
}

\caption{Training efficiency and recommendation performance of HSTU compared with our proposed \model{}. Scaling trends are highlighted with fitted lines for both methods.}\label{fig:scaling}

\end{figure}
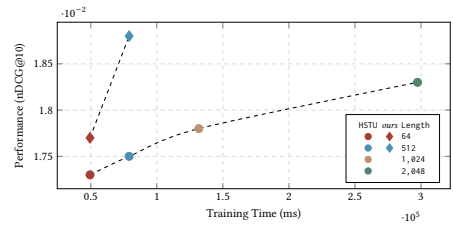

Specifically, recent advances have explored how GR performance scales with data volume~\cite{wukong}, sequence length~\cite{longer}, network depth~\cite{hstu} and hidden dimension~\cite{fuxi}. 
While these findings highlight the potential of scaling as a pathway to stronger recommendation performance, practical deployment in real-world systems presents unique challenges. As shown in Figure~\ref{fig:scaling}, although longer sequences can improve recommendation accuracy, they also substantially increase training and inference time as well as GPU memory consumption, making it impractical to model long user histories without efficiency-oriented solutions.


Although some industrial efforts have explored these issues~\cite{onerec, genrank}, existing approaches often involve complex designs or focus primarily on optimizing network architectures rather than addressing the efficiency of user sequence modeling (see a detailed discussion in Sec.~\ref{sec:related}). In this paper, we propose a simple categorical user sequence compression framework (\model{}) that partitions user interactions into a recent sequence and a distant historical sequence. By condensing the long-term history into a compact set of \textit{history tokens}, \model{} effectively incorporates richer sequential information while substantially alleviating the computational burden of training and inference on ultra-long sequences. Based on Figure~\ref{fig:scaling}, \model{} achieves substantial improvements in recommendation performance while maintaining comparable training cost. Our extensive evaluation on two large-scale real-world datasets shows that, compared to the influential HSTU model, our method can achieve up to a sixfold reduction in computational cost. Additionally, it can deliver up to 39\% higher accuracy at a similar computational expense (i.e., equivalent sequence length).

%

\section{Preliminaries and Related Work}
\label{sec:related}

\begin{figure*}[]
  \centering
  \includegraphics[width=.65\linewidth]{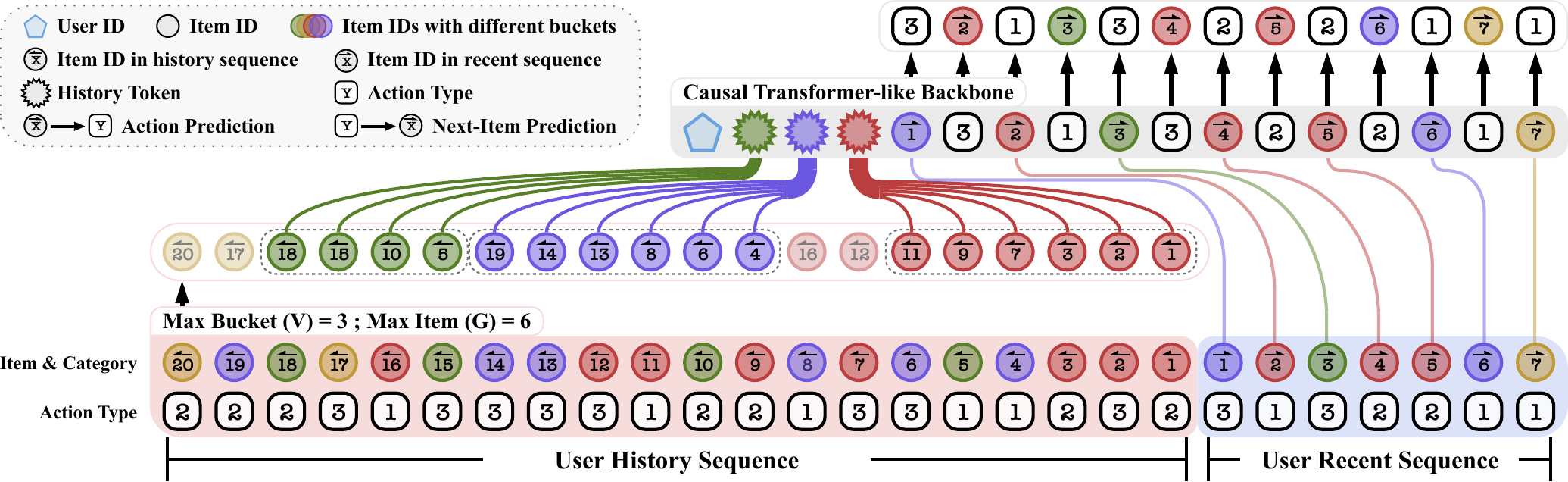}
  \caption{Illustration of the proposed user sequence compression framework. The observed user sequence $\mathbf{s}_u^{\leq t}$ is divided into a long-term history $\mathbf{h}_u$ (in red background) and a recent sequence $\mathbf{r}_u^{\leq t}$ (in blue background). The history is substantially compressed based on item categorical features, while the recent sequence, which reflects the user’s current interests, is used to train both next-item prediction and action prediction tasks.}
  \label{fig:framework}
\end{figure*}

\subsection{Generative Recommendation}


\noindent\textbf{Problem Statement.} Let $\mathcal{U}$, $\mathcal{I}$, and $\mathcal{A}$ denote the sets of users, items, and action types (e.g., click, buy, dislike), respectively, with $|\mathcal{U}| = M$, $|\mathcal{I}| = N$, and $|\mathcal{A}| = T$. 
Each training sample corresponds to a user $u \in \mathcal{U}$ and consists of user observed sequence
$\mathbf{s}_u = [(i_1, a_1), \ldots, (i_L, a_L)]$ of length $L$, where $i_t \in \mathcal{I}$, $a_t \in \mathcal{A}$.

The generative recommendation task  aims to model the conditional probability of the next interaction based on the observed sequence $\mathbf{s}_u^{\leq t}$:
\begin{equation}
    P(i_{t+1}, a_{t+1} \mid \mathbf{s}_u^{\leq t}, u).
\end{equation}

\noindent\textbf{Training Tasks.} Two distinct training objectives are employed.

\textit{Next-item prediction.}
A contrastive InfoNCE loss with negative sampling is utilized for next-item prediction. For the $t$-th position embedding $\mathbf{e}'_t \in \mathbb{R}^d$, the ground-truth next-item embedding $\mathbf{e}^{(i)}_{\text{pos}}$, and $K$ sampled negatives $\{\mathbf{e}^{(i)}_k\}_{k=1}^K$, the probability of predicting the correct item is
\begin{equation}
P(i_{\text{pos}} | \mathbf{e}'_t) = \frac{\exp(\mathbf{e}_t^{\prime\top} \mathbf{e}^{(i)}_{\text{pos}} / \tau)}{
\exp(\mathbf{e}_t^{\prime\top} \mathbf{e}^{(i)}_{\text{pos}} / \tau) + \sum_{k=1}^{K} \exp(\mathbf{e}_t^{\prime\top} \mathbf{e}^{(i)}_k / \tau)},
\end{equation}
with temperature $\tau > 0$. The loss is the negative log-likelihood:
\[
\mathcal{L}_{\text{item}} = - \log P(i_{\text{pos}} | \mathbf{e}'_t).
\]

\textit{Action prediction (optional).}
Given $\mathbf{e}'_t$ and an action vocabulary $\mathcal{A}$, a softmax classifier is used for action prediction:
\[
\hat{\mathbf{y}}^{(a)}_t = \mathrm{softmax}(\mathbf{W}_a \mathbf{e}'_t + \mathbf{b}_a),
\]
where $\mathbf{W}_a \in \mathbb{R}^{|\mathcal{A}| \times d}$, $\mathbf{b}_a \in \mathbb{R}^{|\mathcal{A}|}$. 
With ground-truth label $a_t \in \mathcal{A}$, the cross-entropy loss is
\[
\mathcal{L}_{\text{action}} = - \log \hat{\mathbf{y}}^{(a)}_t[a_t].
\]
\vspace{-0.7cm}
\textbox{\textbf{Opportunities}: Recent studies show that scaling laws hold w.r.t. user sequence length~\cite{longer} and model parameter size~\cite{wukong}.}

\subsection{Efficiency Challenge}

Despite the promise of generative approaches and scaling laws, recommender systems pose unique challenges in real-time responsiveness and computational cost~\cite{liu2024benchmarking,liu2023fans,liu2025can}. The stringent latency requirements in online scenarios place practical limitations on how large or deep generative models can scale. Consider a decoder-only causal Transformer with depth $H$, sequence length $L$, hidden dimension $D$, and feed-forward network (FFN) expansion $r$. Ignoring constant factors, the computational cost per layer is
$\mathcal{O}(L^2 D) + \mathcal{O}(L D^2 r)$,
corresponding to the self-attention and FFN components, respectively.
For $H$ layers, the total cost is $\mathcal{O}(H L^2 D) + \mathcal{O}(H L D^2 r).$

In recommendation settings, the embedding dimension $D$ is typically small (e.g., $64$), while the user sequence length $L$ can be large, resulting in the attention term becoming the primary computational bottleneck $\text{Cost} \approx \mathcal{O}(H L^2 D).$
\textbox{\textbf{Challenges}: Increasing user sequence length leads to quadratic growth in both training and inference costs.
}



Several studies have highlighted the efficiency bottlenecks of long-sequence modeling in industrial settings. For example, in click-through rate prediction, TwinV2~\cite{twin} from Kuaishou employs hierarchical clustering to generate sequence representatives, while LONGER~\cite{longer} from ByteDance introduces a token mixer to fuse neighboring items. In the context of generative recommendation, GenRank~\cite{genrank} from Xiaohongshu reduces sequence length in HSTU by aggregating each item with its corresponding action rather than compressing user behaviors. More recently, OneRec~\cite{onerec} from Kuaishou adopted a multi-stage pipeline that combines clustering (as in TwinV2) with Q-Former for historical sequence compression. However, the complexity of this approach, coupled with the absence of open-source implementation, poses significant challenges for reproducibility, comprehensive evaluation, and further development. In this work, we propose a simple yet effective method for user sequence compression that leverages inherent item categorical features. Our approach offers a practical solution for significantly accelerating generative recommender systems. Further, our open-source codebase facilitates reproducibility and enables the research community to easily build upon and extend our work.


\begin{table*}[!t]

\centering

\renewcommand{\arraystretch}{1.0} 
\caption{Main Results. ``iLen'' and ``sLen'' represent user item sequence length, and input sequence length to the backbone model, respectively. The columns ``Inference'' and ``Training'' correspond to the experimental runtime, measured in milliseconds (ms) and seconds (s), respectively. The "Inference" column reports the average time required to compute a single batch, while the "Training" column presents the time taken to train for one epoch. We use ``N@k'' to represent NDCG metric.}
\label{tab:big-table}
\scalebox{0.85}{
\begin{tabular}{clllcccccccc}
\toprule
\textbf{Exp} & \textbf{Model} & \textbf{iLen} & \textbf{sLen} & \textbf{N@1} & \textbf{N@10} & \textbf{N@20} & \textbf{N@100} & \textbf{N@200} & \textbf{MRR} & \textbf{Inference (ms)} & \textbf{Training (s)} \\
\midrule
\multicolumn{12}{c}{\cellcolor{bglight} \textbf{Dataset:} Kuairand-27K} \\
\midrule
\texttt{(a)} & HSTU & 64 & 128 & 0.0061 & 0.0173 & 0.0218 & 0.0356 & 0.0427 & 0.0163 & 0.21 & 49.41 \\
\texttt{(b)} & $\quad$+\model{} & 64 & 128 & 0.0061 & 0.0177 & 0.0228 & 0.0369 & 0.0442 & 0.0169 & 0.19 & 45.73 \\
\texttt{(c)} & HSTU & 256 & 512 & 0.0060 & 0.0173 & 0.0220 & 0.0362 & 0.0435 & 0.0165 & 0.84 & 57.88 \\
\texttt{(d)} & HSTU & 512 & 1,024 & 0.0061 & 0.0175 & 0.0221 & 0.0364 & 0.0437 & 0.0167 & 1.90 & 79.04 \\
\texttt{(e)} & HSTU & 1,024 & 2,048 & 0.0060 & 0.0178 & 0.0227 & 0.0370 & 0.0442 & 0.0169 & 4.39 & 131.80 \\
\texttt{(f)} & HSTU & 2,048 & 4,096 & \underline{0.0063} & \underline{0.0183} & \underline{0.0233} & \underline{0.0383} & \underline{0.0459} & \underline{0.0174} & 11.10 & 297.35 \\
\texttt{(g)} & $\quad$+\model{} & 512 & 1,024 & \textbf{0.0068} & \textbf{0.0188} & \textbf{0.0237} & \textbf{0.0384} & \textbf{0.0459} & \textbf{0.0179} & 1.87 & 79.19 \\
\multicolumn{2}{r}{Impr. \texttt{(g)} w.r.t. \texttt{(d)}} & 512 & 1,024 & 11.48\% & 7.43\% & 7.24\% & 5.49\% & 5.03\% & 7.19\% & 1$\times$ & 1$\times$ \\
\multicolumn{2}{r}{Impr. \texttt{(g)} w.r.t. \texttt{(f)}} & 2,048 & 4,096 & 7.94\% & 2.73\% & 1.72\% & 0.26\% & 0.00\% & 2.87\% & \textbf{6$\times$} & \textbf{4$\times$} \\
\midrule
\texttt{(h)} & GenRank & 512 & 512 & 0.0038 & 0.0121 & 0.0157 & 0.0266 & 0.0325 & 0.0117 & 1.56 & 64.94 \\
\texttt{(i)} & $\quad$+\model{} & 512 & 512 & 0.0041 & 0.0126 & 0.0163 & 0.0277 & 0.0339 & 0.0122 & 1.47 & 61.47 \\
\multicolumn{2}{r}{Impr. \texttt{(h)} w.r.t. \texttt{(i)}} & 512 & 512 & 7.89\% & 4.12\% & 3.82\% & 4.14\% & 4.31\% & 4.27\% & 1$\times$ & 1$\times$\\
\midrule[1pt]
\multicolumn{12}{c}{\cellcolor{bglight} \textbf{Dataset:} MovieLens-20M} \\
\midrule
\texttt{(j)} & HSTU & 64 & 128 & \underline{0.0315} & 0.0965 & 0.1202 & 0.1710 & 0.1886 & 0.0840 & 0.18 & 23.58 \\
\texttt{(k)} & $\quad$+\model{} & 64 & 128 & 0.0404 & 0.1168 & 0.1421 & 0.1928 & 0.2095 & 0.1006 & 0.18 & 24.19 \\
\texttt{(l)} & HSTU & 128 & 256 & 0.0302 & 0.0987 & 0.1233 & 0.1756 & 0.1930 & 0.0852 & 0.24 & 31.45 \\
\texttt{(m)} & HSTU & 256 & 512 & 0.0304 & 0.1004 & 0.1259 & 0.1791 & 0.1962 & 0.0864 & 0.41 & 54.26 \\
\texttt{(n)} & HSTU & 512 & 1,024 & 0.0312 & \underline{0.1030} & \underline{0.1283} & \underline{0.1819} & \underline{0.1991} & \underline{0.0883} & 0.89 & 114.96 \\
\texttt{(o)} & $\quad$+\model{} & 256 & 512 & \textbf{0.0423} & \textbf{0.1249} & \textbf{0.1519} & \textbf{0.2045} & \textbf{0.2207} & \textbf{0.1066} & 0.42 & 55.11 \\
\multicolumn{2}{r}{Impr. \texttt{(o)} w.r.t. \texttt{(m)}} & 256 & 512 & 39.14\% & 24.40\% & 20.65\% & 14.18\% & 12.48\% & 23.38\% & 1$\times$ & 1$\times$ \\
\multicolumn{2}{r}{Impr. \texttt{(o)} w.r.t. \texttt{(n)}} & 512 & 1024 & 35.58\% & 21.26\% & 18.39\% & 12.42\% & 10.85\% & 20.72\% & \textbf{2}$\times$ & \textbf{2}$\times$ \\
 \midrule
\texttt{(p)} & GenRank (f) & 64 & 64 & 0.0244 & 0.0748 & 0.0938 & 0.1402 & 0.1582 & 0.0664 & 0.15 & 19.08 \\
\texttt{(q)} & $\quad$+\model{} & 64 & 64 & 0.0289 & 0.0854 & 0.1067 & 0.1549 & 0.1730 & 0.0754 & 0.16 & 20.29 \\
\multicolumn{2}{r}{Impr. \texttt{(p)} w.r.t. \texttt{(q)}} & 64 & 64 & 18.44\% & 14.17\% & 13.75\% & 10.49\% & 9.36\% & 13.55\% & 1$\times$ & 1$\times$ \\
\bottomrule
\end{tabular}
}
\end{table*}

\section{Categorical User Sequence Compression}

\textbox{\textbf{Our Solution}: Condense user histories into compact tokens to retain past user interests and reduce computational costs.
}

To address the efficiency challenges posed by long user sequences, we propose a categorical user sequence compression (\model{}) framework. The core idea is to separate user observed sequence $\mathbf{s}_\mathbf{u}^{\leq t}$ into a long-term history $\mathbf{h}_u$ and a recent sequence $\mathbf{r}_\mathbf{u}^{\leq t}$, i.e., $\mathbf{s}_\mathbf{u}^{\leq t} = (\mathbf{h}_u; \mathbf{r}_\mathbf{u}^{\leq t})$. The history will be compressed into a compact set of history tokens, i.e., $\model{}\left(\mathbf{h}_u\right)$, thereby preserving past user interests while substantially reducing training and inference costs. Therefore, we formulate the generative recommendation as:
\begin{equation}
P(i_{t+1}, a_{t+1} \mid \mathbf{r}_u^{\leq t}, \model{}\left(\mathbf{h}_u\right), u).
\end{equation}
Here, we propose a simple yet effective method that leverages item categorical information for user sequence compression.


Item categories serve as the core features in our \model{} algorithm, used to assign items into buckets, from which compressed history tokens are subsequently aggregated. 
They can either be directly obtained from raw metadata or constructed from post-trained representations (e.g., applying K-Means~\cite{kmeans} over learned item embeddings).
\model{} supports both one-to-many (each item is associated with a unique category) and many-to-many (each item may belong to multiple categories) mappings: an item is assigned to every categories it belongs to.

\textbf{\textit{Step 1: Grouping.}}
Let $c \in \mathcal{C}$ be a categorical feature associated with items, where $|\mathcal{C}| = V^C$. 
We group user history sequence $h_u^+$ into at most $V$ buckets, one per category, where $V \leq V^C$.

\textbf{\textit{Step 2: Inter-Bucket and Inner-Bucket Selection.}}
Buckets are first ordered according to the recency of user interactions, determined by the timestamp of the most recent interaction within each bucket. From these, only the top-$V$ most recent buckets are selected and concatenated as input, yielding a compressed history token sequence of length $V$.
Within each selected bucket, up to $G$ of the most recent items are preserved, also sorted by their timestamps. Figure~\ref{fig:framework} illustrates an example with $V=3$ and $G=6$, where the earliest \textcolor{brown!80!black}{\textbf{brown}} bucket is excluded, and the 12$^{th}$ and 16$^{th}$ items in the \textcolor{red!80!black}{\textbf{red}} bucket are discarded.

\textbf{\textit{Step 3: Aggregation.}} 
For each selected bucket $j$, we derive a bucket-level history embedding $\mathbf{e}^{(h)}_j$ by aggregating the embeddings of items assigned to this bucket:
\begin{equation}
\mathbf{e}^{(h)}_j = \frac{1}{|B_j|} \sum_{i \in B_j} \left( W_{\text{align}} \mathbf{E}^{(i)}[i] + b_{\text{align}} \right) + \mathbf{E}^{(b)}[j],
\end{equation}
where $B_j (|B_j| \leq G)$ denotes the set of items grouped into the $j$-th bucket, $\mathbf{E}^{(i)}[i] \in \mathbb{R}^d$ is the embedding of item $i$, $\mathbf{E}^{(b)}[j] \in \mathbb{R}^d$ represents the $j$-th bucket embedding, and $W_{\text{align}} \in \mathbb{R}^{d' \times d}$ with $b_{\text{align}} \in \mathbb{R}^{d'}$ are trainable parameters that project item embeddings into the bucket-aligned semantic space. Finally, the compressed history is formed by concatenating the bucket-level embeddings:
\[
\mathbf{E}^{(h)} = [\mathbf{e}^{(h)}_1, \mathbf{e}^{(h)}_2, \ldots, \mathbf{e}^{(h)}_v].
\]
These history tokens are then included in the model input to represent long-term preferences.

\noindent\textbf{Input Sequence Construction.}
Our input sequence consists of three segments: the user ID, the compressed historical tokens, and the recent user sequence. Each segment is separated by a special token at the beginning. Following the design of HSTU~\cite{hstu}, we flatten the user sequence into alternating item and action tokens, as demonstrated in Figure~\ref{fig:framework}. For each item token, its embedding is integrated with the corresponding categorical feature.

\noindent\textbf{Model Training.} The total training loss is $\mathcal{L} = \mathcal{L}_{\text{item}} + \mathcal{L}_{\text{action}}.$


\section{Experiments}

\subsection{Experimental Setup}

\noindent \textbf{Data Pre-processing.}  
Dataset statistics are provided in Table~\ref{tab:statistics}.

\textit{\textbf{KuaiRand-27K}} consists of user interaction logs collected over 31 consecutive days in a micro-video recommendation scenario. Data from days 1–15 is treated as long-term user history.
Data from days 16–23 is used for training generative recommendation models, while day 24 is reserved for validation and testing. 
To mitigate label imbalance, only two action types i.e., \textit{click} and \textit{long view}, are considered during training. \textit{\textbf{MovieLens-20M}} is a widely used benchmark for movie recommendation, containing user–item interactions spanning several decades, with 5 action types. We follow the preprocessing protocol of Fuxi-$\alpha$~\cite{fuxi}.

\noindent\textbf{Implementation Details.}  
Positive interactions are used to construct both the user history and recent sequences.
Models are trained for 200 epochs, and the checkpoint with the best validation performance is reported on the test set. Latency (inference and training time) is measured on a single idle GPU device. The negative sampling ratio is fixed at 200. The maximum number of buckets $V$ is set to 8, and the maximum number of items per buckets $G$ is set to 32. We employ the Adam optimizer, tuning the learning rate from \{1e-3, 2e-3, 5e-3, 1e-2\} and selecting 5e-3 as the optimal value. For a fair comparison across methods, the number of Transformer layers is fixed to 3, the hidden dimension to 64, and the number of attention heads to 8. We evaluate performance using the nDCG@k (shortly N@k) and MRR metrics, with $k \in \{1, 10, 20, 100, 200\}$.

\begin{table}[t]
\caption{Dataset statistics. ``iLen'' represents the length of user recent sequence.}\label{tab:statistics}
\centering

\setlength\tabcolsep{3pt}

\resizebox{.8\linewidth}{!}{
\begin{tabular}{l|ccccc|cccc}
\toprule
\textbf{\cellcolor{bglight} Dataset} & \multicolumn{5}{l|}{\cellcolor{bglight} \textbf{Kuairand-27K}} & \multicolumn{4}{l}{\cellcolor{bglight} \textbf{MovieLens-20M}} \\
\textit{Description} & \multicolumn{5}{l|}{\textit{220K Items; 27K Users}} & \multicolumn{4}{l}{\textit{21K Items; 130K Users}} \\
\textit{Category} & \multicolumn{5}{l|}{\textit{MusicType (one-to-many)}} &  \multicolumn{4}{l}{\textit{Genres (many-to-many)}} \\
\midrule
iLen & 64 & 256 & 512 & 1,024 & 2,048 & 64 & 128 & 256 & 512 \\
\midrule
\# Train & 228K & 68K & 41K & 30K & 27K & 130K & 130K & 130K & 130K \\
avg (history) & 1,968 & 1,706 & 1,389 & 877 & 327 & 35 & 20 & 8 & 2 \\
avg (recent) & 64 & 252 & 488 & 848 & 1,267 & 41 & 57 & 68 & 74 \\

\bottomrule
\end{tabular}
}

\end{table}

\subsection{Main Results}

Table~\ref{tab:big-table} presents the effectiveness and efficiency of the proposed \model{} across two recommendation datasets, from which we observe the following:
\textbf{A. Scaling law in sequence length.} Recommendation quality consistently improves as the sequence length increases, accompanied by rising computational costs, as also illustrated in Figure~\ref{fig:scaling}.  
\textbf{B. Comparison with HSTU and GenRank.} While GenRank reduces efficiency cost relative to HSTU, it suffers from unacceptable performance degradation. For example, on KuaiRand-27K, GenRank ($\textit{iLen}=512$) is approximately $1\times$ faster than HSTU with the same length, but its efficiency is comparable to HSTU at $\textit{iLen}=256$, with N@10 dropping substantially from 0.0173 to 0.0121.  
\textbf{C. Overall superiority of \model{}.} Across both datasets and backbone architectures, \model{} is consistently lighter and more effective. Relative to the same sequence length, \model{} achieves up to 39.14\% improvements. Even when compared with the strongest baselines trained on longer sequences, \model{} attains superior accuracy with dramatically lower computational cost. Importantly, \model{} is model-agnostic, supporting both HSTU and GenRank backbones, thereby demonstrating strong scalability.



\begin{table}[t]
\caption{Performance under missing vs. inherent item category features on the \textbf{Kuairand-27K} dataset (iLen=512).}
\label{tab:kmeans}

\resizebox{.8\linewidth}{!}{
\begin{tabular}{llllll}
\toprule
\textbf{History} & \textbf{Item Feature} & \textbf{N@1} & \textbf{N@10} & \textbf{N@100} & \textbf{MRR} \\
\midrule
MusicType & MusicType & 0.0060 & 0.0178 & 0.0370 & 0.0169 \\
\midrule
N/A & N/A & 0.0042 & 0.0127 & 0.0288 & 0.0125 \\
K-Means & N/A & 0.0045 & 0.0135 & 0.0297 & 0.0131 \\
K-Means & K-Means & 0.0048 & 0.0144 & 0.0325 & 0.0140 \\
\bottomrule
\end{tabular}
}
\end{table}



\begin{table}[t]
\caption{Ablation Studies.
}
\label{tab:ablation}

\resizebox{.75\linewidth}{!}{
\begin{tabular}{lllll}
\toprule
 & \textbf{N@1} & \textbf{N@10} & \textbf{N@100} &  \textbf{MRR} \\
 \midrule
 \multicolumn{5}{c}{\cellcolor{bglight} \textbf{Dataset:} Kuairand-27K (iLen=512)} \\
 \midrule
\model{}$_\text{HSTU}$ & \textbf{0.0068} & \textbf{0.0188} & \textbf{0.0384} & \textbf{0.0179} \\
 $\quad$w/o $W_\text{align} \& b_\text{align}$ & 0.0064 & 0.0180 & 0.0378 & 0.0172 \\
 $\quad$w/o \textit{action prediction} & 0.0058 & 0.0178 & 0.0375 & 0.0168 \\
 $\quad$w/o \textit{history} $\mathbf{h}_\text{u}$ & 0.0061 & 0.0175 & 0.0364 & 0.0167 \\
\midrule
\multicolumn{5}{c}{\cellcolor{bglight} \textbf{Dataset:} MovieLens-20M (iLen=64)} \\
\midrule
\model{}$_\text{HSTU}$ & \textbf{0.0404} & \textbf{0.1168} & \textbf{0.1928} & \textbf{0.1006} \\
 $\quad$w/o $W_\text{align} \& b_\text{align}$ & 0.0395 & 0.1136 & 0.1897 &  0.0982 \\
 $\quad$w/o \textit{action prediction} & 0.0406 & 0.1144 & 0.1896 & 0.0988\\
 $\quad$w/o \textit{history} $\mathbf{h}_\text{u}$ & 0.0315 & 0.0965 & 0.1710 & 0.0840 \\
\bottomrule
\end{tabular}
}
\end{table}


\subsection{Ablation Studies}

As shown in Table~\ref{tab:kmeans}, when inherent item categories are unavailable, we cluster items via K-Means~\cite{kmeans} and use cluster centers as categorical features. This clustering-based approach improves performance and further benefits when applied to both history compression and recent sequences, though it remains inferior to using inherent categories. These results show that \model{} can be extended to scenarios without explicit category information.

Moreover, as shown in Table~\ref{tab:ablation}, we study the effect of the reminder component proposed in our method.
From the results, we find that removing either the alignment weights, action prediction, or history leads to a noticeable drop in performance, indicating that each contributes positively to the model's predictive ability.

\section{Conclusion}
In this paper, we address the efficiency challenge in generative recommendation by introducing a simple yet effective categorical user sequence compression method, which significantly reduces sequence length and computational cost while enhancing recommendation performance. This approach offers practical value for deploying efficient generative recommendation in real-world systems. We hope our method and codebase will facilitate further research and enable large-scale generative recommendation.

\bibliographystyle{ACM-Reference-Format}
\balance
\bibliography{sample-base}

\end{document}